\documentclass[graybox]{svmult}
\usepackage{helvet}         		
\usepackage{courier}        		
\usepackage{type1cm}        	
\usepackage{makeidx}         	
\usepackage{graphicx}        	
\usepackage{multicol}        	
\usepackage[bottom]{footmisc}	
\makeindex   
\begin{document}          			
\title*{Classical aspects of Hawking radiation verified in analogue gravity experiment.}
\authorrunning{Measurement of stimulated Hawking radiation} 
\author{Silke Weinfurtner, Edmund W.~Tedford, Matthew C.~J.~Penrice, William G.~Unruh, and Gregory A.~Lawrence}
\institute{Silke Weinfurtner \at SISSA - International School for Advanced Studies
via Bonomea 265, 34136 Trieste, Italy, \email{silkiest@gmail.com}
\and Edmund W.~Tedford \at Department of Civil Engineering, University of British Columbia, 6250 Applied Science Lane, Vancouver, Canada V6T 1Z4.  \email{name@email.address}
\and Matthew C. J. Penrice \at Department of Physics and Astronomy, University of Victoria, Victoria, Canada V8W 3P6 \email{mattpen@uvic.ca}
\and William G.~Unruh \at Department of Physics and Astronomy, University of British Columbia, Vancouver, Canada V6T 1Z1 \email{unruh@physics.ubc.ca}
\and Gregory A.~Lawrence \at Department of Civil Engineering, University of British Columbia, 6250 Applied Science Lane, Vancouver, Canada V6T 1Z4. \email{lawrence@civil.ubc.ca}}
%
%
\maketitle
\abstract{
There is an analogy between the propagation of fields on a curved spacetime and shallow water waves in an open channel flow. By placing a streamlined obstacle into an open channel flow we create a region of high velocity over the obstacle that can include wave horizons. Long (shallow water) waves propagating upstream towards this region are blocked and converted into short (deep water) waves. This is the analogue of the stimulated Hawking emission by a {white hole} (the time inverse of a black hole). The measurements of amplitudes of the converted waves demonstrate that they appear in pairs and are classically correlated; the spectra of the conversion process is described by a {Boltzmann-distribution}; and the Boltzmann-distribution is determined by the determined by the change in flow across the white hole horizon.}
\section{Motivation} \label{sec:analoguegravity}
There is a broad class of systems where perturbations propagate on an effective $(d+1)$ dimensional spacetime geometry. In the literature this phenomenon is referred to as an \emph{analogue model}. The first modern paper on analogue spacetime geometry was published in 1981 by W.G.~Unruh~\cite{Unruh:1981bi}, followed by Matt Visser~\cite{Visser:1993tk} in 1993. It was demonstrated that \emph{sound waves} in a fluid flow propagate along geodesics of an \emph{acoustic} spacetime metric. 
More generally, for a single scalar field $\phi$ whose dynamics is governed by some generic Lagrangian $\mathcal{L}(\partial_a \phi, \phi)$, the kinematics of small perturbations around some background solution,
$\phi(t,\vec{x}) = \phi_{0} (t,\vec{x}) + \epsilon \, \phi_{1} (t, \vec{x}) + \frac{\epsilon^{2}}{2} \, \phi_{2}(t, \vec{x}) + \dots $,
can be described by a minimally coupled free scalar field,
$\left( \Delta _{g(\phi_{0})} - V(\phi_{0}) \right) \phi_{1} = 0$, where $\Delta _{g(\phi_{0})}$, a d'Alembertian operator with metric tensor
\begin{equation}
g_{ab}(\phi_{0}) = \left[ - \det \left( \frac{\partial^{2}  \mathcal{L}}{\partial(\partial_{a} \phi) \, \partial (\partial_{b}  \phi) } \right)  \right]^{\frac{1}{d-1}} \Bigg\vert_{\phi_{0}} \;  \;
\left( \frac{\partial^{2}  \mathcal{L}}{\partial(\partial_{a} \phi) \, \partial (\partial_{b}  \phi) } \right)^{-1} \Bigg\vert_{\phi_{0}} \, , 
\end{equation}
an effective curved spacetime geometry~\cite{Barcelo:2002dp}. Over the last 25 years the basic concept of analogue models has been transferred to many different media, and by now we know of a broad class of systems that possess an effective spacetime metric tensor as seen by linear excitations. Detailed background information and current developments can be found in~\cite{Barcelo:2005ln}.

\begin{figure}[h!]
\includegraphics[width=1\textwidth]{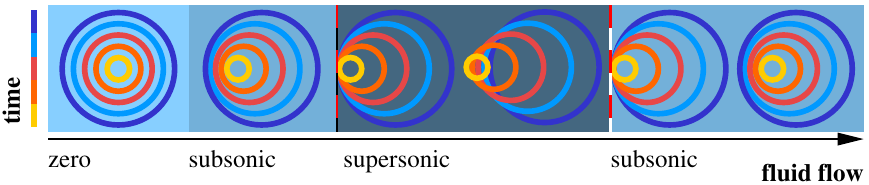}
\caption{Acoustic horizons. The propagation of sound waves in a convergent fluid flow exhibiting sub- and supersonic flow regions are depicted. The dashed (red) black\,/\,white lines, separating the sub- and supersonic regions, indicate the location of the acoustic black\,/\,white horizon. From the left to the right the flow velocity is speeding up and slowing down again.}
\label{FIG:acousticBH}       
\end{figure}

Analogue models of gravity provide not only a theoretical but also an experimental framework in which to verify predictions of classical and quantum field theory in curved spacetimes. For example, the first model, proposed by W.G.~Unruh in 1981, is based on the fact that sound waves propagating on an inviscid and irrotational fluid flow satisfy the Klein--Gordon equation in an effective curved background~\cite{Unruh:1981bi}. If the velocity of the fluid exceeds the velocity of sound at some closed surface, a dumb hole, i.e. an analogue of a black hole, forms, see Fig.~\ref{FIG:acousticBH}. The presence of effective horizons opens up new possibilities to experimentally explore the black hole evaporation\,/\,Hawking radiation process. 

There are several hinderances that one has to overcome before testing analogue gravity systems in a laboratory experiment. Any experimental setup has to fall within the approximations made when deriving the analogy. For example, the main difficulty in implementing acoustic black hole (dumb hole) horizons, is to ensure that the waves obey the linear approximation throughout. Shock waves (sonic booms) occur far to readily at transitions between sub- and supersonic flows. In fact, we are all familiar with the sonic boom related to the shock wave generated by supersonic aircrafts. 

In 2002 it was argued that {surface waves} in an open channel flow with varying depth are an ideal toy model for black hole experiments~\cite{Schutzhold:2002di}. Unruh's 1981 paper raised the possibility of doing experiments with these analogues. One issue with Hawking's derivation is its apparent reliance on arbitrarily high frequencies, this phenomenon is commonly referred to as the {trans-Planckian problem}\footnote{The original derivation by Hawking radiation predicts that the quantum field excitations in the initial state --- which are responsible for the late time radiation --- have frequencies exponentially higher than the frequency associated with the Planck scale~\cite{Brout:1995aa,Jacobson:1999aa}.}. The dispersion relation of gravity waves creates a natural physical short wavelength cutoff, which obviates this difficulty. Thus the dependence of the Hawking effect on the high-frequency behavior of the theory can be tested in such analogue experiments~\cite{Unruh:1995ec,Unruh:1995aa,Corley:1996aa}. While numerical studies indicate that the effect is independent of short-wavelength physics, experimental verification of this would strengthen our faith in the process. The presence of this effect in our physical system, which exhibits turbulence, viscosity, and non-linearities, would indicate the generic nature of the Hawking thermal process. Below we present all the necessary steps to understand and carry out such an experiment.

\section{Black \& white hole evaporation process} \label{sec:bhevaporations}
One of the most striking findings of general relativity is the prediction of {black holes,} accessible regions of no escape surrounded by an event horizon. In the early 70s, {Hawking suggested that black holes evaporate via a quantum instability}~\cite{Hawking:1974rv,Unruh:1976aa}. The study of classical and quantum fields around black holes shows that small classical as well as quantum field excitations are being amplified. In particular, a pair of field excitations at temporal frequency $f$ are created, with amplitudes $\alpha_f$, $\beta_f$ (Bogoliubov coefficients) related by,
\begin{equation} \label{Eq:Hawking}
\frac{\vert \beta_f \vert^2}{\vert \alpha_f \vert^2}= \exp \left( \frac{-4\pi^2 f}{g_\mathrm{H}} \right)
\end{equation}
where $g_\mathrm{H}$ is the surface gravity of the black hole, and $\alpha_f$ and $\beta_f$ are positive and negative
norm components~\cite{Hawking:1974rv,Unruh:1976aa}. Positive norm modes are emitted, while negative ones are absorbed by the black hole, effectively reducing its mass. The surface gravity for a non-rotating black hole with a mass $M$ is given by $g_\mathrm{H}=1.0\times10^{35}/M$ [kg/s]. Equation~(\ref{Eq:Hawking}) is applicable for both {stimulated} and spontaneous emission, and at regimes where the quantum physics is dominant. A comparison of~(\ref{Eq:Hawking}) with the {Boltzmann-distribution} allows one to associate a temperature $T$ with the black hole,
\begin{equation}
T=\frac{\hbar \, g_H}{2 \pi k_B} = 1.2 \cdot 10^{-12} \cdot g_\mathrm{H} \, [\mbox{sK}] = 6.03 \cdot 10^{-8} \frac{M_\circ}{M} [\mbox{K}] \, .
\end{equation}
Here $M_\circ$ is a solar mass, and the smallest observed black holes are of this order. Thus black hole evaporation is clearly difficult to observe directly~\cite{Carr:2005lr}.  \\

The situation is not as challenging in an analogue gravity experiment, where one is dealing with table-top experiments that are under much better control and significantly easier to access. The question then arises as to how to collect conclusive experimental evidence to be assured one is dealing with analogue black hole evaporation, and not with some other classical or quantum process. As we will demonstrate below, the Hawking process exhibits in principle the following measurable characteristics: (i) the emission of field excitations is correlated; (ii) the spectra of the emission process is described by a {Boltzmann-distribution}; (iii) the Boltzmann-distribution is determined by the {surface gravity} at the effective horizon; and (iv) the emitted field excitations are stronger-than-classically\,/\,quantum correlated. In the following we will present an analogue gravity experiment in which we observe all classical features of the Hawking process, i.e.~(i)--(iii). We will later argue that it is not practical to look for (iv) due to the particular analogue system we are using.

\section{Experimental setup}
\label{sec:expsetup}
\begin{figure}[h!]
\sidecaption
\includegraphics[scale=1]{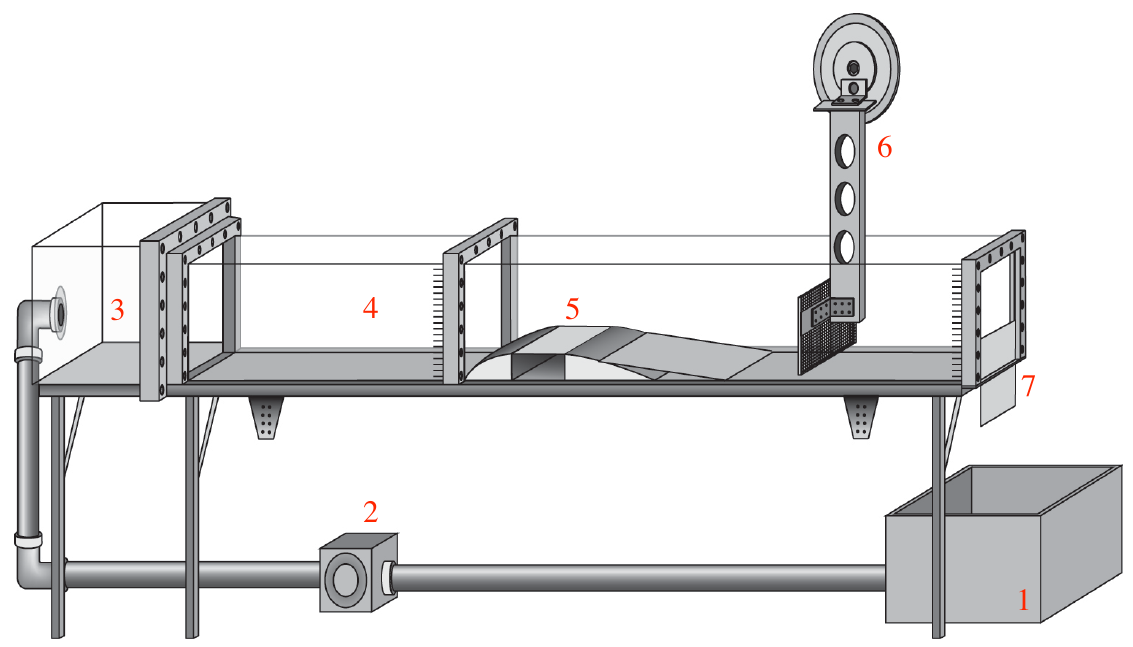}
\caption{Experimental apparatus. The experimental apparatus used in our experiments: 
(1) holding reservoir, 
(2) pump and pump valve,
(3) intake reservoir, 
(4) flume, 
(5) obstacle, 
(6) wave generator, and
(7) adjustable weir.}
\label{FIG:wavetank}       
\end{figure}

Our experiments were performed in a 6.2 m long, 0.15 m wide and 0.48 m deep flume~(Fig.~\ref{FIG:wavetank}), and were partly motivated by experiments in similar flumes~\cite{Badulin:1983fk,Rousseaux:2008uq,Lawrence:1987qy}. We set up a spatially varying background flow by placing a 1.55 m long and 0.106 m high obstacle in the flume.

\begin{figure}[b!]
\sidecaption
\includegraphics[width=7.5cm]{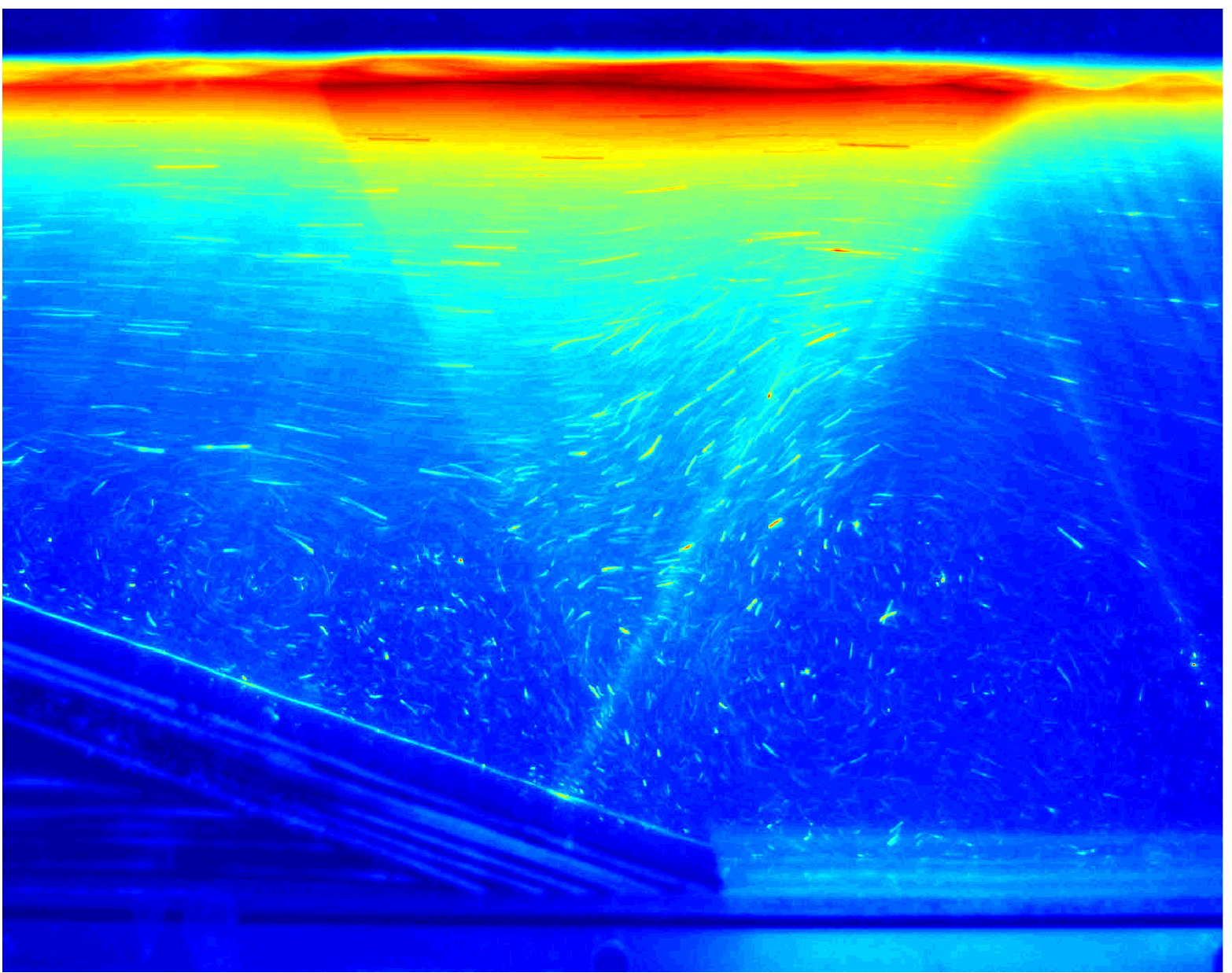}
\caption{Image of flow separation. The image visualize the flow behavior at the lee side of an obstacle with a trapezoidal profile. The visualization technique utilizes emerged neutrally buoyant particles. The motion of the particles during the exposure time causes streak lines indicating approximately the velocity field of the flow.\label{Fig:particlestreaks}}
\end{figure}

\begin{figure*}[ht]
\begin{center}
\leavevmode
\includegraphics[width=1\textwidth]{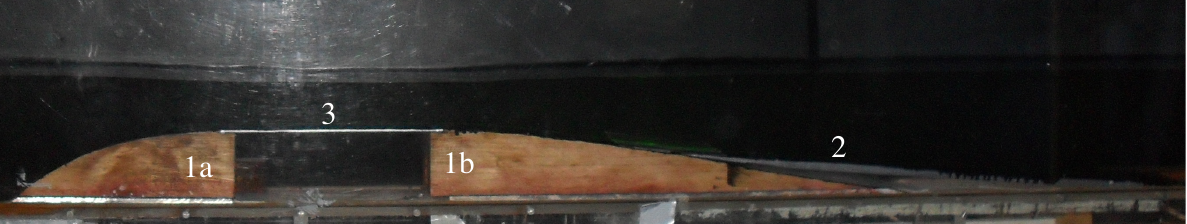}
\caption[Experimental setup diagram.]{Obstacle used for experiments: (1a) and (1b) curved parts motivated by airplane wing; (2) flat aluminum plate to further reduce flow separation; and (3) flat top aluminum plate to reduce wave tunneling effects.}
\label{Fig:experimentalsetup}
\vspace{3mm}
\end{center}
\end{figure*}

Particular care was taken to design an obstacle to minimize, or avoid, flow separation. Especially downstream of the obstacle as the flow slows down it has the tendency to separate and create a recirculating flow, see Fig.~\ref{Fig:particlestreaks}.
Initially our obstacle was modeled after an airplane wing with a flat top and a maximum downstream slope of 5.2 degrees designed to prevent flow separation, with a profile given by
\begin{equation}\label{obst}
H(x) = 2a\Bigl(1-x-\exp(-b \ x)\Bigr),
\end{equation}
where $a=0.094$ of a meter and $b=5.94$ per meter. However the gradual change in slope along the down stream side of the profile, as well as the absence of any sharp transitions, were not sufficient to fully prevent flow separation. To address this issue we added a constant slope along the backside of the obstacle. The plate is $0.81$ meters long. It tapers at a $4.5$ degree angle on each end so as to create a smooth transition from obstacle to plate and then from the plate to the bottom of the flume. The gradual slope eliminated apparent flow separation. Maximum flow velocity occurs at the crest of the obstacle. In order to reduce wave tunneling effects between the effective black and white hole horizons, the crest of the obstacle was extended. This was done by cutting the obstacle at the crest and adding a plate ($15$ cm in length) to join the sections. The extended flat section at the crest of the obstacle resulted in a region of relatively uniform maximum flow velocity. The final obstacle is displayed in Fig.~\ref{Fig:experimentalsetup}.
%
%

We used particle imaging velocimetry~\cite{Adrian:2005rt} to determine the flow rate $q$, and to verify the suppression of flow separation. In this technique, small neutrally buoyant, tracer particles  are added to the fluid. A short light pulse from each of two lasers (with different colors, red and green, say) is focussed into a narrow sheet
within the fluid,  the two pulses being separated in time by a few milliseconds. The light scattered in a direction normal to the sheet by the tracer particles within the sheet is focussed so that it forms an image of the particles in a monochrome CCD camera.  Each particle produces two images, separated by a distance that is a measure of the component of the velocity in the plane of the light sheet with which the particle is moving; the distance between the two images is therefore a measure of the component of the local fluid velocity.  Analysis requires that the pair of images belonging to a given particle be identified, and this is achieved by a cross-correlation technique based on the assumption that particles within a small interrogation area are moving with approximately the same velocity. As a result one obtains the flow velocity as a function of height $v(h)$, and the flow rate is given by $q=\int_0^{h_0} v(h) \, dh$.

Shallow water waves of approximately 2 mm amplitude were generated 2 m downstream of the obstacle, by a vertically oscillating mesh, which partially blocked the flow as it moved in and out of the water. The intake reservoir had 
flow straighteners and conditioners to dissipate turbulence, inhomogeneous flow, and surface waves caused by the inflow from the pump. The flume was transparent to allow photography through the walls, and the experimental area was covered to exclude exterior light.

We are interested in the excitations propagating on the background flow in our setup. We measured and analyzed the variations in water surface height using essentially the same techniques as in~\cite{Tedford:2009fk}. The water surface was illuminated using laser-induced fluorescence, and photographed with a high-resolution (1080p) monochrome camera. The fluoresces served to  scatter the light to the sides where it could be photographed, to sharply delineate the surface, since the mean free path of the laser in the dyed water was less than $1\,\mbox{mm}$, and to suppressed the speckle which bedevils all laser illuminated objects.

\begin{figure}[h!]
\sidecaption
\includegraphics[width=7.5cm]{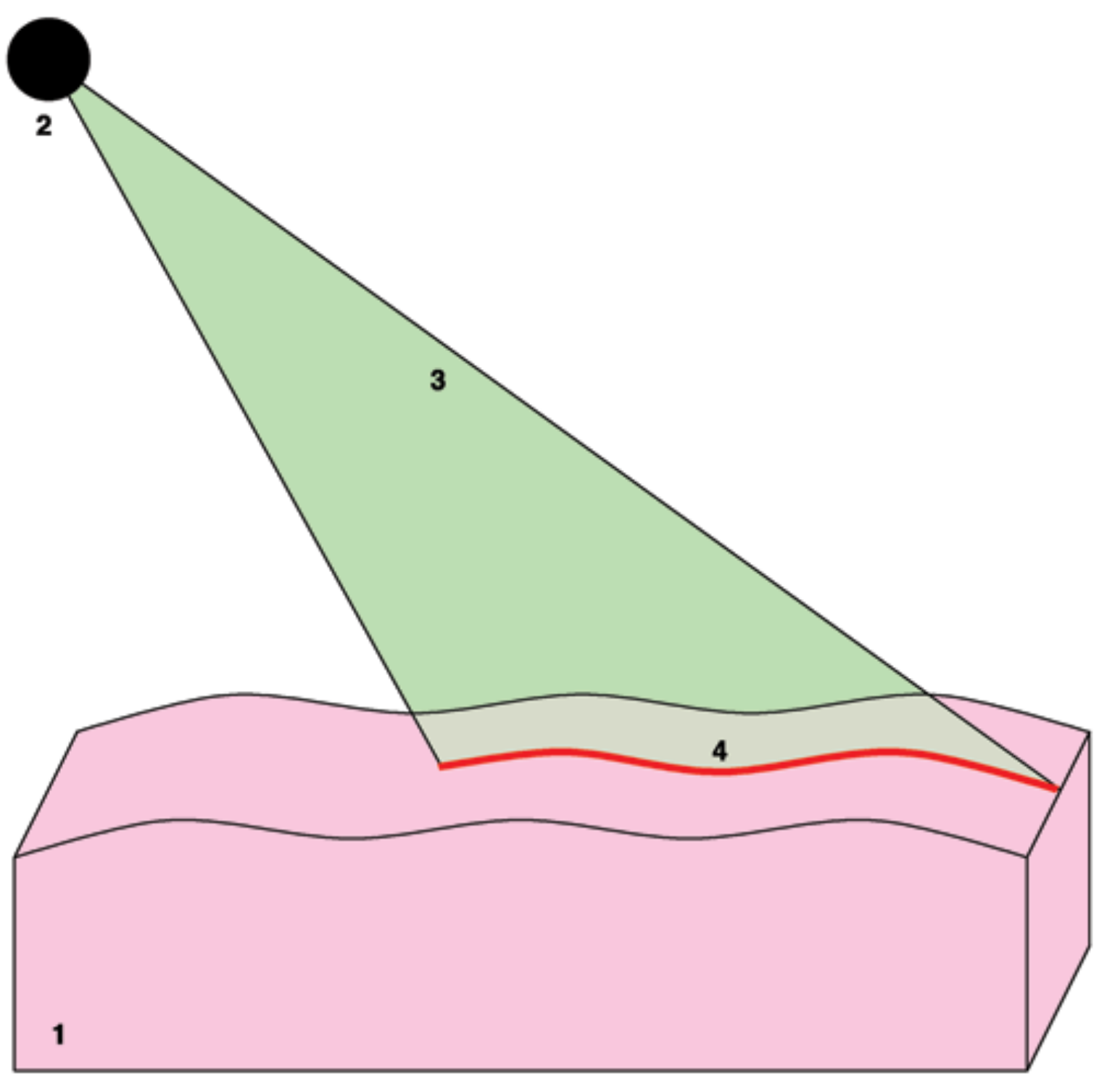}
\caption{Surface wave detection. Diagram of light-sheet projection for surface wave detection: (1) water with dye, (2) Powell lens, (3) light sheet, and (4) fluorescing water surface.}
\label{FIG:lightsheet}       
\end{figure}

The camera was set up such that the pixel size was 1.3 mm, the imaged area was 2 m wide and 0.3 m high, and the sampling rate was 20 Hz. The green (532 nm) 0.5 W laser light passed through a Powell lens to c reate a thin ($\sim$ 2 mm) light sheet (Fig.~\ref{FIG:lightsheet}). Rhodamine-WT dye was dissolved in the water, which fluoresced to create a sharp ($<$ 0.2 mm) surface maximum in the light intensity. We interpolated the intensity of light between neighboring pixels to determine the height of the water surface to subpixel accuracy. 

\section{Quasi-particle excitations}
\label{sec:excitations}
The {excitation spectrum} of gravity waves on a slowly varying background flow is well understood and has a dispersion relation given by,
\begin{equation}
f^2=\left( \frac{g \, k}{2 \pi} \right) \cdot \tanh \left( 2\pi \, k \, h \right) \, ,
\end{equation}
with the frequency, $f = 1/\omega$, where  is the wave period; the wavenumber, where $k=1/\lambda$ is the wavelength; $g$ is the gravitational acceleration, and h the depth of the fluid. We neglect surface tension and viscosity. We classify waves according to the value of $2\pi \, k\,h$. Our waves all had wavelengths longer than about $2.1\,\mbox{m}$ (still water wavelengths), and surface tension would only play a role for waves with wavelengths less than about $1\,\mbox{cm}$.

For  $2\pi \, k\,h<1$ the dispersion relation can be approximated by $f = \sqrt{gh} k$. These shallow water waves (called that because their wavelength $1/k$ is much longer than the depth of the water $h$) have both group and phase speed approximately equal to $\sqrt{gh}$. For $2\pi\,k\,h > 1$, the dispersion relation is approximated by $f = (gk/2\pi)^{1/2}$. The group speed of these deep water waves is approximately half the phase speed, and both vary as the square root of the wavelength. For a given water depth, both the group and phase speeds of deep water waves are less than the group and phase speeds of shallow water waves. 

\begin{figure}[h!]
\sidecaption
\includegraphics[width=7.5cm]{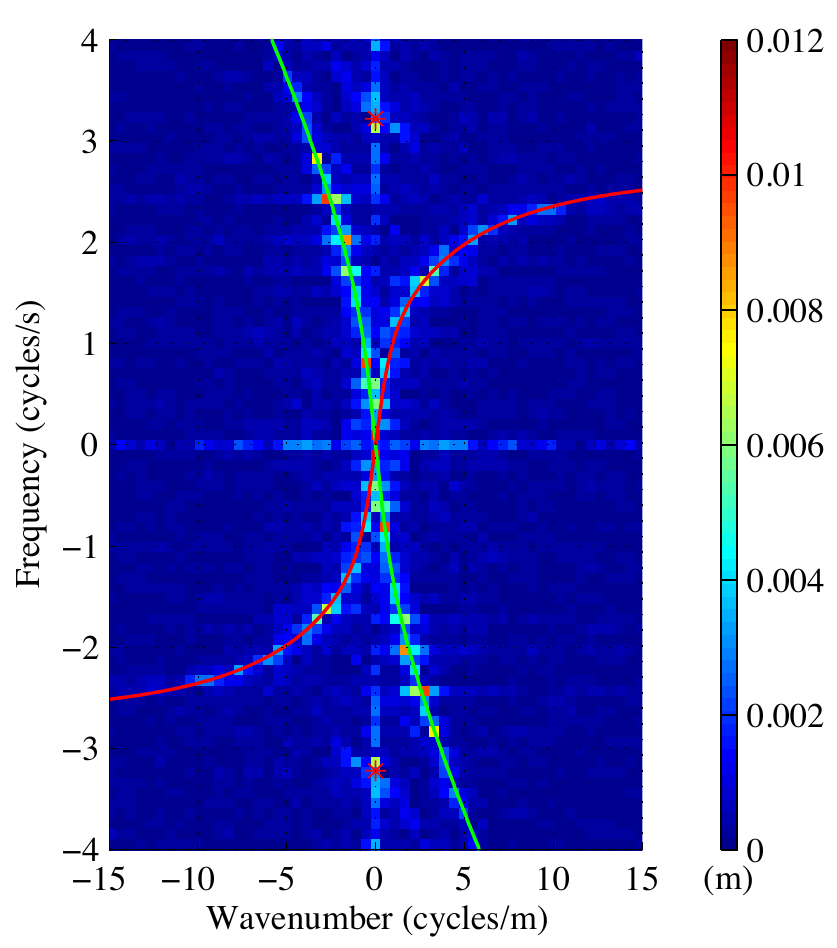} %
\caption[Background noise.]{Background noise. Fourier transform of water surface in flat bottom flume without waves; $q=0.039\,\mbox{m}^2/\mbox{s}$ and $h=0.24\,\mbox{m}$. Fluctuations lie on upstream (red line) and downstream (green line) branches of the dispersion relation. Just visible at $f=\pm3.1$, $k=0$ are the second transverse mode branches of the dispersion relation. Off the dispersion curves, the background noise amplitudes are less than $0.1\,\mbox{mm}$. }
\label{FIG:BackgroundNoise}       
\end{figure}

To determine the ambient wave noise in our facility, and to check the effectiveness of our procedures, we conducted an experiment without the obstacle in place and with no wave generation. The space and time Fourier transform of the noise match the dispersion relation for this flow ($q=0.039\, \mbox{m}^2/\mbox{s}$ and $h=0.24\, \mbox{m}$) extremely well (Fig.~\ref{FIG:BackgroundNoise}). In general, the amplitude of the Fourier components has a noise level of less than 0.2 mm away from the dispersion curves. The apparently elevated noise energy crossing the k axis at $f=\pm3.1\, \mbox{Hz}$ is due to the second transverse mode branch of the dispersion relation (the first transverse mode has a node at the location of the light sheet).

\begin{figure}[h!]
\sidecaption
\includegraphics[width=7.5cm]{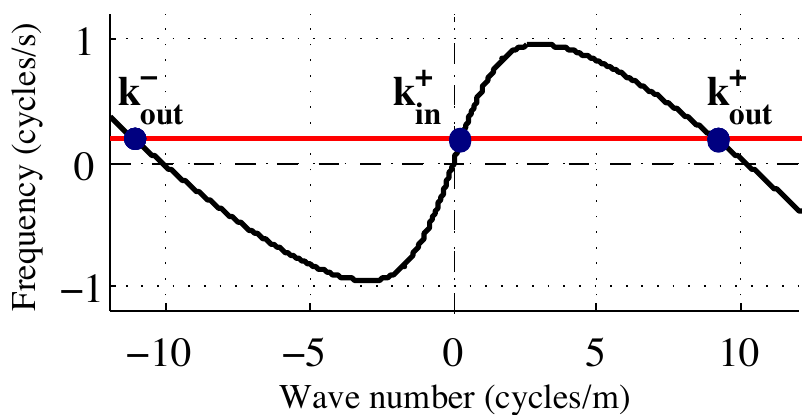} %
\caption{Conversion process. Dispersion relation for waves propagating against a flow typical of our experiments. A shallow water wave, $k_{in}$, sent upstream, is blocked by the flow and converted to a pair of deep water waves ($k^+_{out}$ and $k^-_{out}$) that are swept downstream.}
\label{FIG:Conversion}       
\end{figure}

In~\cite{Schutzhold:2002di} SchŸtzhold and Unruh argued that the equation of motion of shallow water waves can be cast into a wave equation on a curved spacetime background if the speed of the background flow varies. Assuming a steady, incompressible flow the velocity
\begin{equation}
v(x)=\frac{q}{h(x)}.
\end{equation}
Here the two-dimensional flow rate q is fixed. The dispersion relation in the presence of a non-zero background velocity becomes,
\begin{equation}
(f+v\,k)^2=\left( \frac{g \, k}{2 \pi} \right) \cdot \tanh \left( 2\pi \, k \, h \right) \, .
\end{equation}
In Fig.~\ref{FIG:Conversion}, the dispersion relation is plotted for a flow typical of our experiments. Only the branch corresponding to waves propagating against the flow is plotted. For low frequencies, there are three possible waves, which we denote according to wavenumber. The first, $k^+_{in}$, is a shallow water wave with both positive phase and group velocities,
and corresponds to the wave that we generate in our experiments. The second, $k^+_{out}$, has positive phase velocity, but negative group velocity. Both waves, $k^+_{in}$ and $k^+_{out}$, are on the positive norm branch of the dispersion relation. The third, k-out, has both negative phase and group velocities, and it lies on the negative norm branch. In our experiment, generated shallow water waves move into a region where they are blocked by a counter-current, and converted into the other two waves. The goal of our experiment was the measurement of the relative amplitudes of the outgoing positive and negative norm modes to test the validity of~(\ref{Eq:Hawking}). (Further conversion from deep-water waves to capillary waves \cite{Badulin:1983fk,Rousseaux:2010md} are also possible but are not studied here.)
The conversion from shallow water to deep water waves occurs where a counter-current become sufficiently strong to block the upstream propagation of shallow water waves~\cite{Rousseaux:2010md,Suastika:2004uq,Unruh:2008rt}. It is this that creates the analogy with the white hole horizon in general relativity. That is, there is a region that the shallow water waves cannot access, just as light cannot enter a white hole horizon. Note that while our experiment is on white hole horizon analogues, it is because they are equivalent to the time inverse of black hole analogues that we can apply our results to the black hole situation.

\section{Experimental procedure}
\label{sec:expprocedure}
We are interested in the physics around white hole, not black hole horizon. In our particular analogue gravity system the two ``outgoing'' modes are now not on either side of the horizon but both come out downstream of the white hole horizon.
In order to measure the effect of the horizon on incident waves, we sent shallow water waves toward the effective white hole horizon, which sits on the lee side of the obstacle. We conducted a series of experiments, with $q=0.045\,\mbox{m}^2/\mbox{s}$ and $h=0.194\,\mbox{m}$, and examined $9$ different ingoing frequencies between $0.02$ and $0.67\,\mbox{Hz}$, with corresponding still water wavelengths between $2.1$ and $69$ meters, corresponding to $0.67$ to $0.02\,\mbox{Hz}$ frequencies. This surface was imaged at $20$ frames per second, for about $200\,\mbox{s}$. In all cases we analyzed a period of time which was an exact multiple of the period of the ingoing wave, allowing us to carry out sharp temporal frequency filtering of the signals (i.e., eliminating spectral leakage).

The analysis of the surface wave data was facilitated by introducing the convective derivative operator $\partial_t + v(x) \partial_x$. We redefine the spatial coordinate using,
\begin{equation}\label{Eq:xi}
\xi=\int_{x=0} \frac{dx}{v(x)} dx \, ,
\end{equation}
where $x$ is the distance downstream from the right hand edge of the flat portion of the obstacle. The coordinate has dimensions of time, and its associated wave number has units of Hz. The convective derivative becomes $\partial_t + \partial_\xi$, or, in Fourier transform space, $f + \kappa$. This is the term that enters the conserved norm. From equations (35), (36)
and (87) of reference~\cite{Schutzhold:2002di} we find that the conserved norm has the form
\begin{equation}\label{Eq:NewNorm}
\int \frac{\vert A(f,\kappa)\vert^2}{(f+\kappa)} d\kappa \, ,
\end{equation}
where $A(f,\kappa)$ is the $t$, $\xi$ Fourier transform of the vertical displacement of the wave. In using this coordinate system the outgoing waves have an almost uniform wavelength even over the obstacle slope.

\section{Data analysis and results}
\label{sec:results}

\begin{figure*}[ht]
\begin{center}
\leavevmode
\includegraphics{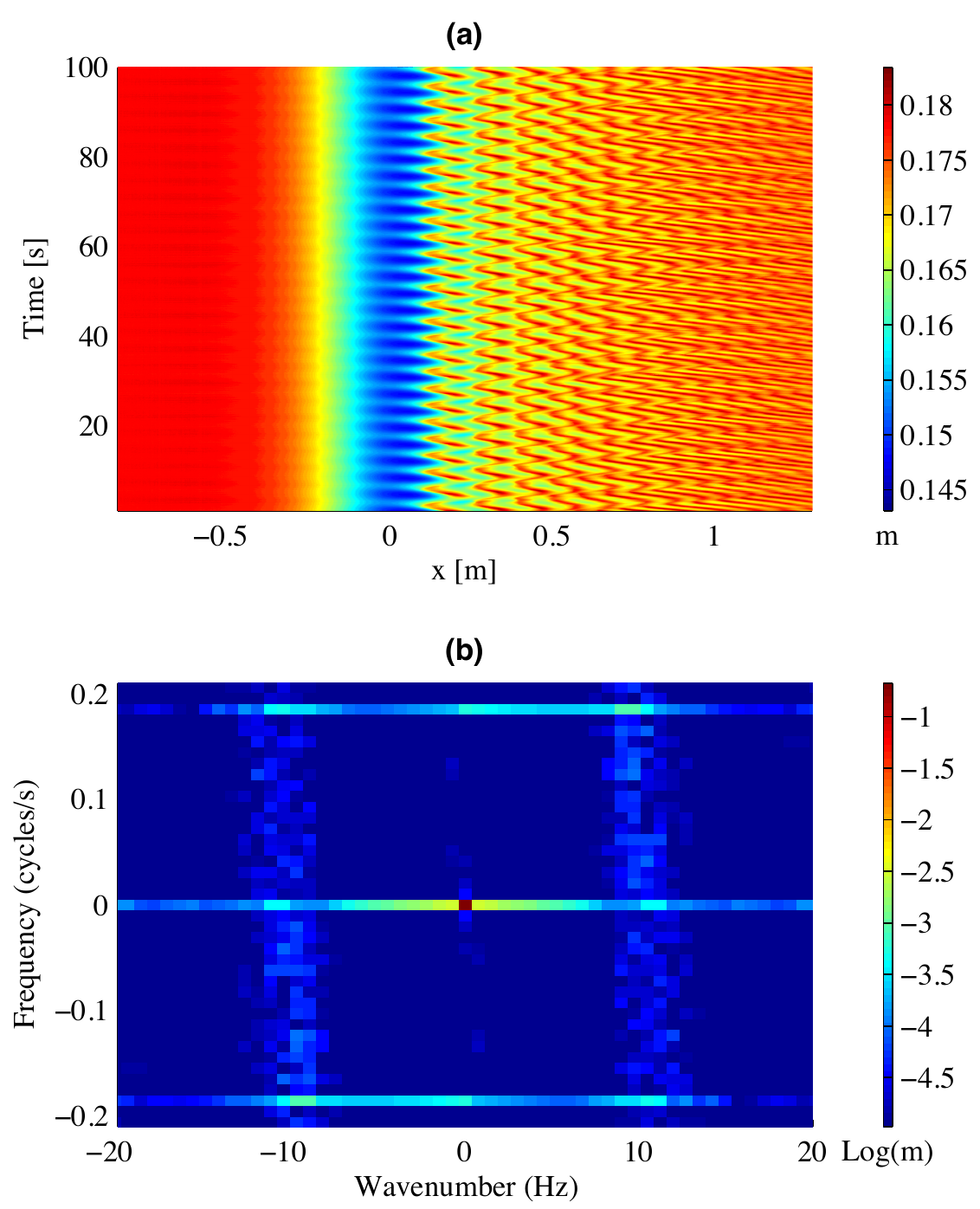}
\caption[Free surface in position (a) and momentum (b) space.]{The wave characteristics (a) shows the changes in the free surface. Notice the colors indicate the water level relative to the bottom of the tank, not that actual water heights. The double Fourier transformation (b) of the wave characteristics shows three excited frequency bands. The one at $\omega=0$ represents the background (at $k=0$) and a standing wave, refereed to as the undulation (to peaks at $k\sim \pm 10\,\mbox{Hz}$). The other two excited frequency bands at $\pm 0.185\,\mbox{cycles}/\mbox{s}$ correspond to the stimulated frequency bands.}
\label{Fig:WCFFT}
\end{center}
\end{figure*}

We will illustrate the pair-wave creation process by presenting the results for $f_{in}=0.185\,\mbox{Hz}$. In this case we analyzed images from exactly 18 cycles, measuring the free surface along approximately $2\,\mbox{m}$ of the flow including the obstacle. We calculated from the wave characteristics, and after converting to $\xi$-coordinates (\ref{Eq:xi}), the two-dimensional Fourier transformation as displayed in Fig.~\ref{Fig:WCFFT} (a) and (b). Note that the amplitudes of the Fourier transform at frequencies above and below $\pm0.185\,\mbox{Hz}$ are very small, indicating that the noise level is small.
%
%

As expected, there are three peaks, one corresponding to the ingoing shallow water wavelength around $\kappa=0$, and the other two corresponding to converted deep water waves peaked near $\kappa_{out}^+=9.7\,\mbox{Hz}$ and $\kappa_{out}^-=-10.5\,\mbox{Hz}$. The former is a positive norm and the latter a negative norm outgoing wave, see equation~(\ref{Eq:NewNorm}).

%
\begin{figure*}[ht]
\begin{center}
\leavevmode
\includegraphics{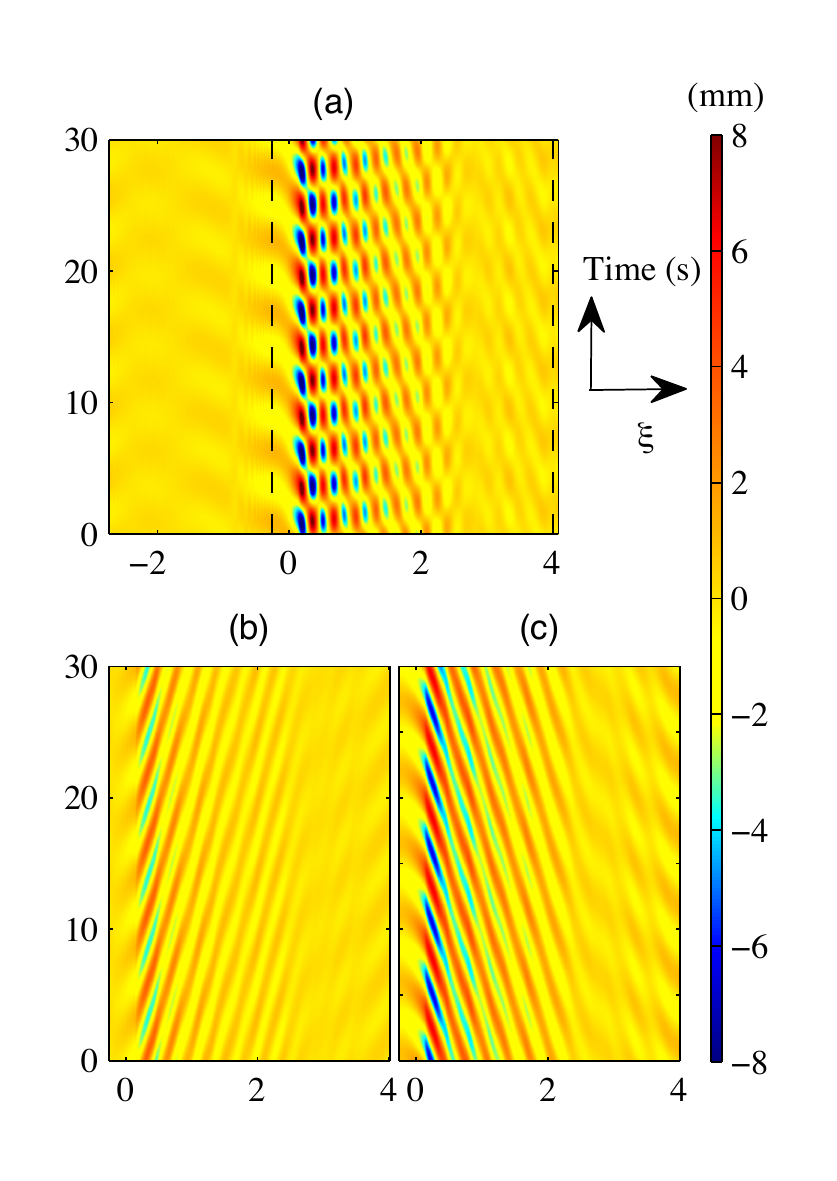}
\caption[Pair-conversion process.]{Demonstration of pair-wave conversion of an ingoing frequency of $0.185\,\mbox{cycles}/\mbox{s}$: a, Fourier transform of unfiltered wave characteristic. b, Filtered wave characteristic containing only the ingoing frequency band. c and d, wave characteristics for filtered negative and positve norm modes. (The colours represent the amplitudes of the waves, see color bars.)}
\label{Fig:FilteredWC}
\end{center}
\end{figure*}
%

In Fig.~\ref{Fig:FilteredWC} we plot the wave characteristics (amplitude as function of $t$ and $\xi$) filtered to give only the temporal $0.185\,\mbox{Hz}$ band. Figures~\ref{Fig:FilteredWC}(b) and \ref{Fig:FilteredWC}(c) are the characteristic plots where we further filter to include only $\kappa<-1\,\mbox{Hz}$ and $\kappa>1\,\mbox{Hz}$ respectively. These are the negative and positive norm outgoing components without the central peak of the ingoing wave (because of their very long wavelengths and the rapid change in wavelength as they ascend the slope, the incoming waves have a very broad Fourier transform). Recall, because we are only interested in counter-propagating waves, we defined positive phase and group speeds as pointing to the left. As expected from the dispersion relationship, see Fig.~\ref{FIG:Conversion}, the negative norm waves have negative phase velocity, while the positive norm waves have positive phase velocity. The complex structure in the characteristics of Fig.~\ref{Fig:FilteredWC}(a) arises because of the interference between the three components, the original ingoing wave, and the positive and negative norm outgoing waves. In Fig.~\ref{Fig:FilteredWC}(b), we see that the ingoing wave is blocked around $\xi=0$, with only a small component penetrating into the region over the top of the obstacle $\xi<0$.

%
\begin{figure*}[ht]
\begin{center}
\leavevmode
\includegraphics{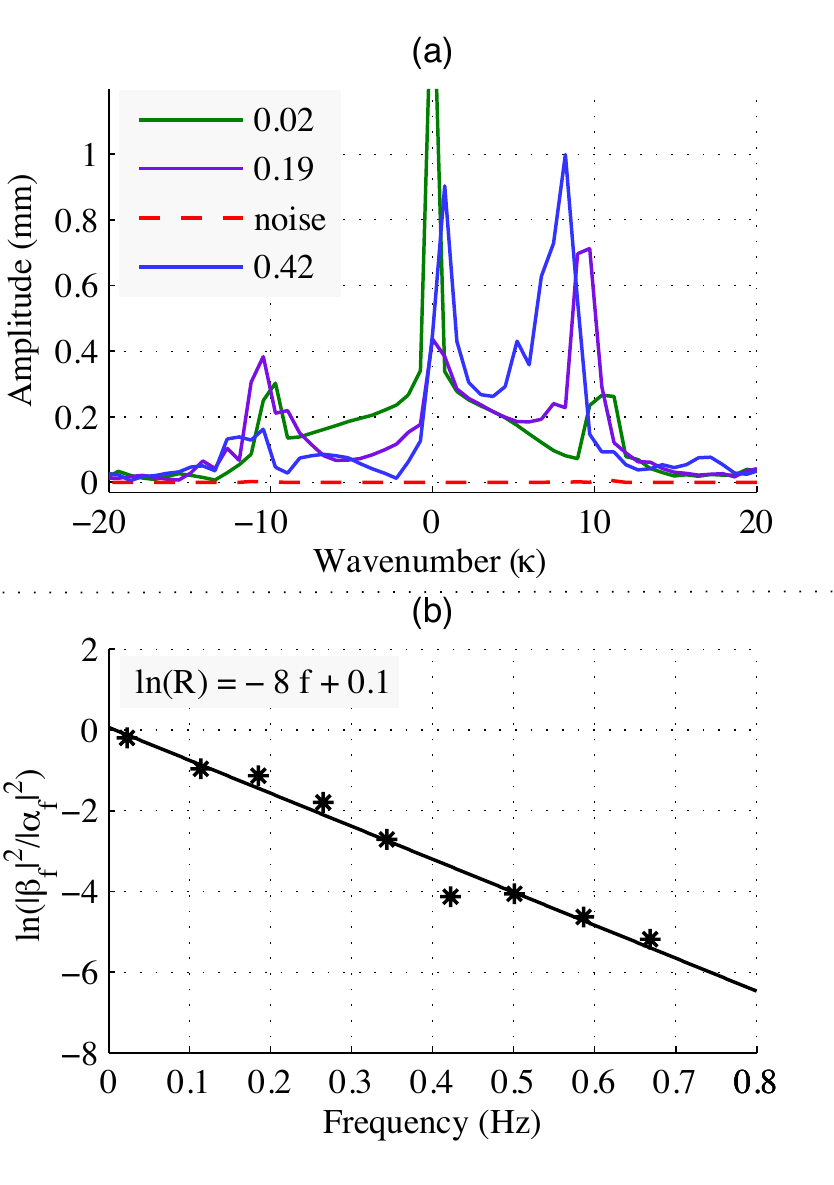}
\caption[Amplitudes and thermal spectrum.]{Amplitudes and thermal spectrum. a, Absolute value of three different ingoing frequency bands, and typical noise level (red line). b, Ratio between negative and positive norm components in between $0.02$ and $0.67\,\mbox{cycles}/\mbox{s}$ (red stars), and linear least-squares fit (red line).}
\label{Fig:Boltzmann}
\end{center}
\end{figure*}
%
Our key results are presented in Fig.~\ref{Fig:Boltzmann}. Figure~\ref{Fig:Boltzmann}(a) shows the amplitude of the spatial Fourier transform at three selected ingoing frequencies. As the frequency increases, the ratio of the negative norm peak to positive norm peak decreases. Furthermore, the location of the positive norm peak moves slightly toward zero as the frequency increases, while the negative norm peak moves away from zero. This is to be expected from the location of the allowed spatial wavenumber from the dispersion plot, see Fig.~\ref{FIG:Conversion}. The red-dashed curve in Fig.~\ref{Fig:Boltzmann}(a) shows the Fourier transform in the adjacent temporal frequency bands for the sample case of $0.185\, \mbox{Hz}$. This is a representation of the noise, and is a factor of at least $10$ lower than the signal in the $0.185$ frequency band.

To test whether or not the negative norm wave creation was due to non-linearities we repeated the runs at all frequencies with 50\% larger amplitudes. The converted wave amplitudes did, in fact, scale linearly.

The crucial question is: Does the ratio of the negative to positive norm outgoing waves scale as predicted by the thermal hypothesis of equation~(\ref{Eq:Hawking})? This is shown to be the case in Fig.~\ref{Fig:Boltzmann}(b), where the norm ratios are plotted as a function of ingoing frequency. To calculate the norm of the outgoing waves we integrate $\int \vert A(f,\kappa)\vert^2 / (f+\kappa) \, d\kappa$ over the peaks. In Fig.~\ref{Fig:Boltzmann}(b) the points represent the log of the ratios of these areas for each of the input frequencies we tested. The thermal hypothesis is strongly supported, with linear regression giving an inverse slope of $0.12\,\mbox{Hz}$ and an offset close to zero. The slope corresponds to a temperature of $T = 6 \times 10^{-12}\, \mbox{K}$, and the offset is zero within our error bounds. 

We see from Fig.~\ref{Fig:FilteredWC}(b,c) that the region of ``wave blocking'' where the ingoing wave is converted to a pair of outgoing waves, is not a phase velocity horizon (where the phase velocity in the laboratory frame goes to zero). This is true even for the very lowest frequencies. The usual derivation of the temperature from the surface gravity relies on this conversion occurring at a phase velocity horizon. This makes the calculation of the surface gravity, and thus the predicted temperature uncertain. In our case estimates of the surface gravity give a predicted temperature of the same order as the measured temperature. What is important is that the conversion process does exhibit the thermal form predicted for the Hawking process.

This, together with the loss of irrotational flow near the horizon, and absence of a dependable theory of surface waves over an uneven bottom make prediction of the temperature from the fluid flow difficult. Our estimates --- using the background flow parameters (i.e.~flow rate and water height) to calculate $g_H=1/2\, \partial(c^2-v^2)/\partial x$ --- give us a value somewhere between about $0.08$ and $0.18\,\mbox{Hz}$. What is important is that the conversion process does exhibit the thermal form predicted for the Hawking process.

\section{Conclusions and outlook}
\label{sec:theend}
We have conducted a series of experiments to verify the stimulated Hawking process at a white hole horizon in a fluid analogue gravity system. These experiments demonstrate that the pair-wave creation is described by a Boltzmann-distribution, indicating that the thermal emission process is a generic phenomenon. It survives fluid-dynamical properties, such as turbulence and viscosity that, while present in our system, are not included when deriving the analogy. The ratio is thermal despite the different dispersion relation from that used by Hawking in his black hole derivation. This increases our trust in the ultraviolet independence of the effect, and our belief that the effect depends only on the low frequency, long wavelength aspects of the physics. When the thermal emission was originally discovered by Hawking, it was believed to be a feature peculiar to black holes. Our experiments, and prior numerical work~\cite{Unruh:1981bi,Corley:1996aa}, demonstrate that this phenomenon seems to be ubiquitous, and not something that relies on quantum gravity or Planck-scale physics.

Black holes are linear phase-insensitive field amplifiers of a very peculiar kind~\cite{Unruh:2011vm,Richartz:2012bd}. As mentioned in the introduction, the energy of the modes suffers an extremely severe de-amplification, going from frequencies and wave numbers far far higher than the Planck scale, to ones in the kHz regime for solar mass black holes. Nevertheless, when looking at the norms of the modes, they act just like any other amplifier. The Hawking effect is the quantum noise which must accompany any amplifier, but the characteristics of that noise are entirely determined by the amplification properties of the amplifier, which can of course be measured in the classical regime. This relation between the classical and quantum behavior was first pointed out by Einstein in his relation between stimulated and spontaneous emission, by Haus and Mullen in their characterization of quantum noise in a linear amplifier, and by many others~\cite{Einstein:1917}. In our case, the direct measurement of the quantum noise, with a characteristic temperature of the order of $T = 6 \times 10^{-12}\, \mbox{K}$, is of course impossible. A possible step forward is to study the behavior of quantum noise in analogue gravity systems in Bose--Einstein condensates, e.g.\cite{Jean-Christophe-Jaskula-LCF:5fk,Kheruntsyan:2012uq}.



\end{document}